# Cross-Lingual News Event Correlation for Stock Market Trend Prediction

Sahar Arshad[1,2],

Nikhar Azhar[1],

Sana Sajid[1],

Seemab Latif[1,†],

Rabia Latif[3]

[1]*School of Electrical Engineering and Computer Science (SEECS), National University of Sciences and Technology (NUST), Islamabad, Pakistan.*

[2]*Department of Computer Science, Bahria University, Islamabad, Pakistan.*

[3]*College of Computer and Information Sciences (CCIS), Prince Sultan University, Riyadh, Saudi Arabia.*

**Keywords**: Natural language-based financial forecasting; Pakistan stock exchange; Financial event timeline; Cross-lingual sentiment analysis

## Abstract

In the modern economic landscape, integrating financial services with Financial Technology (FinTech) has become essential, particularly in stock trend analysis. This study addresses the gap in comprehending financial dynamics across diverse global economies by creating a structured financial dataset and proposing a cross-lingual Natural Language-based Financial Forecasting (NLFF) pipeline for comprehensive financial analysis. Utilizing sentiment analysis, Named Entity Recognition (NER), and semantic textual similarity, we conducted an analytical examination of news articles to extract, map, and visualize financial event timelines, uncovering the correlation between news events and stock market trends. Our method demonstrated a meaningful correlation between stock price movements and cross-linguistic news sentiments, validated by processing two-year cross-lingual news data on two prominent sectors of the Pakistan Stock Exchange. This study offers significant insights into key events, ensuring a substantial decision margin for investors through effective visualization and providing optimal investment opportunities.

[†]Corresponding author: Seemab Latif (Email: seemab.latif@seecs.edu.pk; ORCID: 0000-0002-5801-1568 ).

## 1. Introduction

Financial Technology (FinTech) has significantly reshaped the traditional landscape of the finance industry by introducing innovations and redefining the way financial services are delivered and experienced [1]. The revolutionized adoption of digital financial services is evident through statistics, specifically released in the post-COVID period [2]. According to Servion, within five to ten years from now, 95% of all interactions with clients in the finance industry will be performed by Artificial Intelligence (AI) while only 5% of customer interactions will require human involvement [3]. The financial services roadmap is progressing toward the implementation of proactive, anticipating, and engaging AI assistants in the industry [2].

One important industry in the financial sector is the stock market, where investors participate in trading activities for financial gain. Stock prices are determined by the forces of supply and demand, reflecting investors' perceptions of a company's financial health, growth prospects, and overall market conditions. Other significant factors influencing stock prices include but are not limited to, economic indicators, international transactions, geopolitical events, and market sentiment [4]. Therefore, the market's inherent volatility and unpredictability make it essential for financial advisors to conduct thorough research and stay informed about economic trends and events to strategically manage investment portfolios. In a typical financial environment, investors rely on third-party advisory roles in evaluating market trend statistics to make the most of their financial resources. However, with the digitization of information and real-time chaotic data dissemination, stock market dynamics have become complex to assess for financial advisors and thus for investors to make their timely decisions on the temporal investment opportunities available in the market [5].

By leveraging AI-powered natural language processing (NLP) based financial solutions, Fintech is prevailing in the conventional investment landscapes of financial markets. Investors are now more in charge of their investment strategies, with the utilization of more accessible data analytics and evidence-based accurate calculations [1]. NLP is reshaping FinTech with remarkable achievements in efficiently processing vast amounts of unstructured financial data from diverse sources like news stories and social media updates. Moreover, the useful data insights mitigate the investment risks and enhance the rewards associated with investment strategies. The applicability of the state-of-the-art NLP algorithms along with market historical trends offers robust decision-making, tailored to various types of investors in the financial market. Therefore, reducing the reliance on time and cost-effective consultations from the traditional façade of financial advisors in the industry.

With its evolution and unique capabilities, the research community's tendency to integrate NLP with financial market solutions has been increasing in recent years, thus, contributing to an emerging area of Natural language-based financial forecasting (NLFF) [6]. The intricate NLP techniques like sentiment classification, semantic similarity analysis, and key phrase extraction empower financial analysis from unstructured textual data [7]. The insights from these techniques can be utilized for extracting and visualizing the impact of emotions, news, events, and financial entities in the market trend as well. In conjunction with these techniques, cutting-edge financial forecasting techniques enhance the modeling of the financial market dynamics. This integration proves advantageous for financial analysts and investors, enabling them to anticipate market trends and attain a competitive advantage in the financial realm through informed, data-driven decision-making [8].

Although the inclination towards NLFF within the research community is vibrant with its evolution in recent years, however, it's crucial to understand that financial dynamics, especially within the stock market, are not influenced by geopolitical factors in the same manner across underdeveloped, emerging, and advanced economies. The financial markets in mature economies exhibit notable distinctions from those in underdeveloped Asian-Pacific countries like Pakistan. In the former, indicators such as the issuance of CPI reports, unemployment statistics, and changes in federal reserve interest rates can influence a downturn in their respective financial markets [9]. Conversely, the factors influencing the Pakistan stock market are distinct, with the exchange rate of the dollar, foreign investments, financial

    

reserves, and political pressures serving as major drivers for market movements [10]. Figure 1 illustrates the market trend curve for the year 2022-2023, highlighting the contrast between the S&P 500 Index and the KSE 100 Index for the two representatives from varying economies.

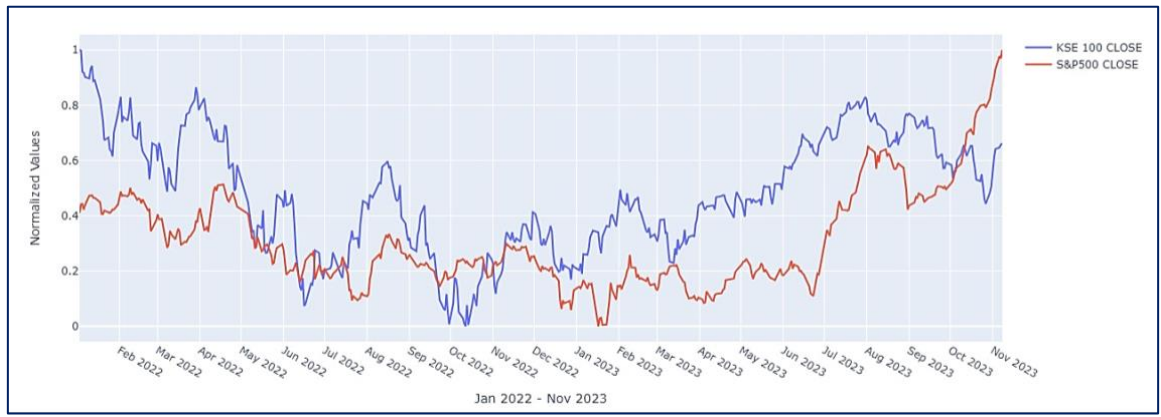

Figure 1: A comparison of the stock trend of the S&P 500 and the KSE 100 Indices

This research aims to provide insights into the unique dynamics of the Pakistan Stock Exchange by focusing on a key component of Natural Language-based Financial Forecasting (NLFF): the generation of financial event timelines through parallel linguistic exploration. The primary focus is to capture and visualize the correlation between events extracted from news sources and the corresponding trends in the stock market. This will enable us to understand the impact of information on economic decision-making and how economic decisions are influenced by sentiments in the economy. The study specifically targets two prominent sectors of the Pakistan Stock Exchange, characterized by higher market capitalization. We have applied the state-of-the-art techniques throughout the research process, leading to multifold contributions:

• Recognizing the distinct dependencies and dynamics of the PSX, a comprehensive dataset that encompasses both Urdu and English news sources is created, allowing for a more precise analysis of local and global factors, correlating each sector's financial landscape.

• Extract, map, and visualize an event's timeline for unstructured news and publicly available stock data for the financial sectors of PSX, leveraging advanced linguistic analysis for reasonable investment decisions in financial modeling techniques. As proof of concept, we applied our event's timeline to analyze the stock market trend analysis, and to a major extent, we were able to map the price movements of stock prices in conjunction with cross-linguistic news sentiments.

The rest of the paper is organized to provide a comprehensive exploration of our research. In Section 2, we delved into the essential components of natural language financial forecasting, emphasizing studies that draw insights from both English and Urdu languages. Section 3 outlines the methodology, providing a detailed account of the approach taken in parallel language data collection, analytical insights, and the development of the event timeline generation model. Results and findings are presented in Section 4, accompanied by a thorough discussion of their implications. We conclude in Section 5, summarizing key insights and proposing avenues for future research.

## 2. Related Work

This section explores key markers of natural language financial forecasting i.e., sentiment analysis, Named Entity Recognition (NER), and event extraction. Emphasis is on investigating their role in providing reasonable data insights and generation of events timeline, to identify the correlation of news events on the stock market tailored to both Urdu and English datasets. At the end of this section, Figure 2 illustrates the taxonomy built through literature study.

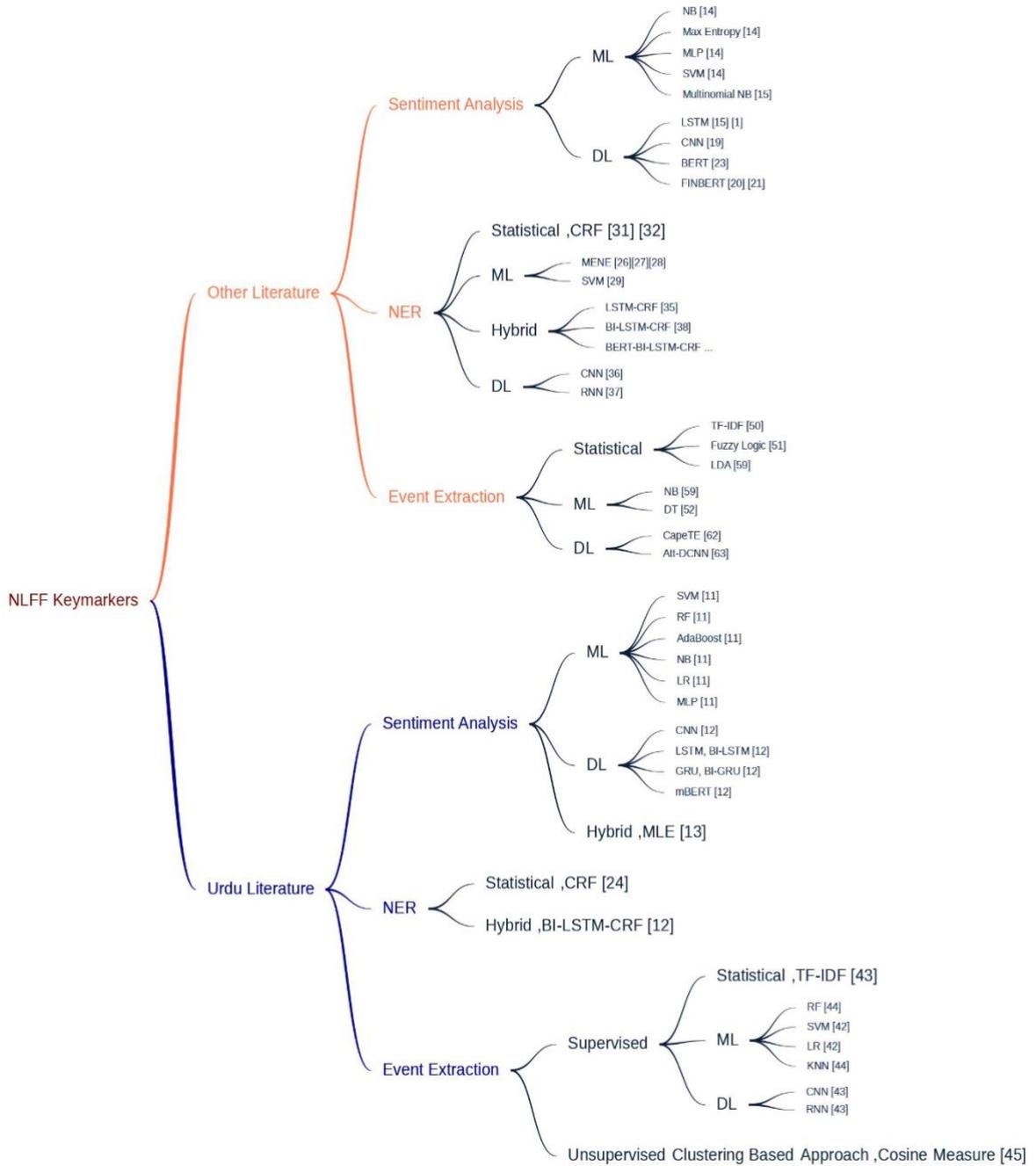

Figure 2. Taxonomy illustration for the cross-lingual domain literature showcasing NLFF key markers.

## 2.1 Financial news sentiment analysis

Sentiment analysis in financial news plays a pivotal role in offering a lens into market sentiments and investor reactions. Its advantages lie in deciphering the emotional tone within financial narratives and providing insights into market perception. Gauging sentiment in NLFF aids in anticipating market trends, assessing risks, and empowering participants with a nuanced understanding of market dynamics and news impact on stock movements to make informed decisions in the future.

### 2.1.1 Exploration of state of the art in urdu language sentiment analysis

The literature on Urdu sentiment analysis presents a comprehensive exploration of various statistical, machine learning, and deep learning models. Khan et al. [11] and [12] contributed significantly by introducing the Urdu Corpus for Sentiment Analysis (UCSA) and UCSA-21 datasets, respectively. They employed a range of machine learning models, including SVM, NB, Adabbost, MLP, LR, RF, and deep learning models like CNN-1D, LSTM, Bi-LSTM, GRU, and Bi-GRU, along with fine-tuned multilingual BERT (mBERT). Results indicate that LR with n-gram feature vectors achieved the highest F1 score at 82.05% for UCSA, while the proposed mBERT classifier achieved 81.49% and 77.18% using UCSA and UCSA-21 datasets, respectively. In another notable study, Masood et al. [16] adopted a combination of CNNs and LSTMs for Urdu sentiment classification, achieving 86% accuracy and 89% F1 score. Ahmed et al. [13] introduced a meta-learning ensemble (MLE) approach, demonstrating its efficacy across benchmark datasets, including SAU-18, UCSA, and UCSA-21. The MLE approach outperforms traditional ensemble methods, showcasing its potential for enhancing deep model performance in Urdu sentiment analysis, with an accuracy of 88.22% for the UCD dataset. This achievement is attributed to classical approaches, given the low-resource nature of Urdu as a language, constrained by limited annotated datasets and pre-trained models. Notably, it underscores the absence of existing financial news data available in the Urdu language. Further refinements in existing approaches can be achieved with contextual sentiment analysis based on ontologies enabling accurate interpretation of sentiments within a specific domain of interest.

### 2.1.2 Exploration of state of the art in financial sentiment analysis

A comprehensive exploration of sentiment analysis in English datasets reveals a progression from classical to modern approaches. Authors in [14] focused on the 2018 Brazilian electoral period and employed Machine Learning techniques such as Naive Bayes, SVM, Maximum Entropy, and MLP for sentiment analysis of tweets and news in Portuguese. While showcasing the multilayer perceptron's superior performance in sentiment analysis and offering insights into the correlation between social media sentiment and stock market movement, the research was limited by its focus on conventional ML approaches, lacking the capability to capture semantic nuances in the text. In [15], authors tackled sentiment analysis for Lithuanian financial texts, by employing machine learning and deep learning models. Multinomial Naive Bayes achieved a higher accuracy of 71.1% followed by LSTM 70.4%. Wu et al. [16] introduced the SI-LSTM approach for stock price prediction and evaluated on China Shanghai A-share market, outperforming traditional machine learning methods. Another study [17] integrated machine learning and LSTM models for accurate stock price prediction, reporting an accuracy of 88.73%. The study enhanced predictions by combining sentiments from news articles along with technical analysis of historical stock data. However, LSTMs struggle with long-range dependencies and highlight a potential constraint in modeling complex relationships compared to newer architectures like Transformers [18]. In another research effort, authors in [19] proposed a Convolutional Neural Network (CNN) for sentiment analysis of Brazilian financial news, which achieved an accuracy of 86.5%. Nevertheless, the challenges linked to CNNs, including their limitations in capturing complex semantics, cannot be ignored. Authors in [20] combined PCA, EMD, and LSTM for stock market price prediction with an achievement of 82% accuracy, showcasing FinBERT's efficacy in sentiment analysis of stock market data of Thailand. However, the application of news sentiment to EMD-LSTM showed inconsistent improvements. The study's limitation lies in its use of traditional or hybrid machine learning models, missing potential advancements in recent model combinations for time series forecasting. [21] presented a two-component model to analyze stock patterns using LSTM and predict stock movements through sentiment analysis with an accuracy of 84.9% on the Kaggle dataset. [22] employed deep learning, particularly LSTM, for stock price prediction, integrating FinBERT-LSTM with superior performance. Although the evaluations of NASDAQ-100 data and New York Times articles demonstrated the model's superior performance from vanilla approaches. Yet, limitations were identified in potential oversight of external factors influencing stock behavior, attributed to the primary focus on news sentiment integration. Lastly, [23] utilized Bidirectional Encoder Representations from Transformers (BERT) for sentiment analysis, achieving a commendable F1 score of 72.5%. Challenges include short data collection periods and interpreting sentiment-analysis-Dow Jones index relationships, prompting future research suggestions. These studies underscore the significance of sentiment analysis while emphasizing the need for advancements to overcome existing limitations and leverage evolving machine learning and NLP techniques specifically in the financial domain.

## 2.2 Financial news named entity recognition

Entity extraction in the financial domain involves identifying and extracting relevant entities such as companies, individuals, and financial instruments from text data. This process is crucial for understanding the relationships and dynamics within the financial market, enabling more accurate analysis and decision-making. Extraction of entities like stock symbols, company names, and key financial indicators, empowers NLFF models, facilitating efficient information acquisition for investors and analysts.

### 2.2.1 Named entity recognition for Urdu language

While exploring the literature on Named Entity Recognition (NER) in Urdu, Wahab Khan et al. [24] introduced a conditional-random-field-based approach for manually annotated Urdu (UNER-I) dataset with seven distinct Named Entity types, showcasing the method's superiority over the baseline techniques. While achieving enhanced F1 scores for both UNER-I and the established IJCNLP-Urdu dataset, the study suggests further improvement by incorporating deep learning techniques for embedding extraction to enhance contextual richness. In another research effort [25], Fida Ullah et al. addressed this challenge through a hybrid methodology. Their proposed Attention-Bi-LSTM-CRF model for NER in Urdu, leveraging the MK-PUCIT Corpus, achieved a remarkable F1-score of 92%, surpassing existing methods and setting a new benchmark for the state of the art in Urdu NER. However, a limitation is noted in the MK-PUCIT Corpus, which included only three NE types, potentially limiting its application in diverse information extraction tasks or inquiries. Notably, limited development of Urdu Named Entity Recognition (UNER) exists due to language diversity and structural uniqueness. Research in this area is scarce, with minimal resources available. Although rule-based methods offer feature design advantages, efforts to contribute to Urdu language resources involve constructing a new homogeneous NE-labeled dataset in the financial domain.

### 2.2.2 Exploration of the state of the art in named entity recognition

This section delves into various models and approaches designed for the NER task, emphasizing the transition from rule-based models to more advanced methodologies. Studies [26-28] presented the Maximum Entropy Named Entity (MENE) based models, to enhance named entity recognition tagging results by leveraging a broad spectrum of knowledge resources. SVM-based SNOOD, proposed by McNamee and Mayfield [29], demonstrated adaptability to various target languages, rendering binary decisions for the current tokens while lacking consideration of neighboring words. CRF-based NER, explored by authors [30-31], overcome this limitation by utilizing latent representations, paving the way for applications in diverse domains including legal [32] and financial domains [33]. However, the limitation lies with the focused domains, potentially hindering the generalizability of the NER task. Various models, such as a multi-tasking deep structural model [34], a dependency-guided LSTM-CRF model [35], Iterative Dilated Convolutional Neural Networks (ID-CNNs) [36], recurrent neural network [37], and a deep learning framework with BiLSTM and CNN [38], were introduced, each addressing distinct aspects of Named Entity Recognition (NER) from multi-task learning to capturing syntactic relationships, addressing lexical features and embeddings in neural network-driven methods. These models demonstrated notable success in applications, including health-related entity identification in Twitter messages, showcasing their effectiveness in handling domain-specific NER tasks. In contrast to the English datasets, significant studies on Chinese datasets [39][40] offered a comprehensive view of recent NER advancements. They covered traditional and deep learning approaches, with [40] introducing a BERT-BiLSTM-CRF model for NER in Chinese medical records, surpassing baselines with an 88.45% F1 score. The collective efforts underscore the continuous evolution of NER methods to address a wide range of challenges in information extraction and understanding.

### 2.3 Financial news event extraction

An event is a particular occurrence having a clear time, location, and involvement of one or more people, frequently denoting a change in status [41]. Event extraction involves identifying and categorizing events mentioned in text data, such as news articles or social media posts. In the context of NLFF, event extraction is crucial for understanding and analyzing events that might impact financial markets. By automatically extracting and categorizing relevant events, NLFF models can leverage this information to make more informed predictions about market movements, helping investors and analysts stay ahead of market trends and make better-informed decisions.

#### 2.3.1 Event extraction for Urdu text

This section explores event extraction in Urdu language literature where multiple studies address diverse aspects, including news classification, deep learning models, multilingual event classification, and document clustering. Study [42] focused on classifying Pakistani news using the Open Data Pakistan dataset, with 97.8% accuracy for Support Vector Machine in single-level classification and 83% for Logistic Regression in multilevel text classification. The study demonstrated the efficacy of machine learning algorithms in news categorization. Another study [43] explored event detection in Urdu language text and achieved 84% accuracy using deep learning models with various feature vectors. While emphasizing the effectiveness of TF-IDF-based vectors, the study acknowledges limitations in pre-trained word embedding models and underscores the importance of addressing dataset imbalance and exploring diverse classification levels for future improvements. In the study [44], the authors pioneered Urdu multi-class event classification at the sentence level for Urdu text from news and social media and achieved high accuracy, 98% with Random Forest and 99% with K-Nearest Neighbor. In a continuing effort for Urdu research, authors in [45] evaluated distance measures for document clustering on Urdu news headlines, highlighting the preference for frequent unigrams with the Cosine distance measure. The findings emphasized the effectiveness of stemming over the lemmatization for news clustering, utilizing a dataset of 1750 Urdu news headlines from popular news agencies. It is evident from the studies that event extraction in Urdu text faces hurdles due to the ongoing evolution and limited availability of annotated data in Urdu literature. Large Language Models (LLMs) like ChatGPT presented a potential solution through simple prompts, eliminating the need for fine-tuning specific datasets. However, experiments conducted in the study [46] revealed that while ChatGPT excels in tasks such as translation and summarization, it encounters challenges in event extraction. The model required intricate instructions defining event types and schemas, as reflected in its 51.04% performance compared to specialized models like Event Extraction with Question-Answering (EEQA) in complex scenarios. This underscores the persistent issue of limited dataset availability and emphasizes the ongoing need for refinement in event extraction tasks.

#### 2.3.2 Study on state of the art in event extraction

The objective of event extraction is to identify key event features within the text and structurally express the events. The empirical investigations carried out by [47] demonstrate that news events can alter investor perceptions, prompt trade, and influence stock price movements. The method is increasingly applied in commercial and financial domains [48], enabling businesses to identify market reactions and derive trading signals. Conventional text filtering for the events involved manual and automatic techniques [49]. Manual approaches require extensive knowledge bases and rule sets, limiting their adaptability to specific domains, whereas automatic methods, such as TF-IDF [50], may underperform on specialized texts. Knowledge heuristics, including fuzzy logic [51], showed promising results when incorporated into advanced statistical strategies but needs manual rule generation. Supervised solutions such as ML-based approaches [52] and information extraction from pre-trained language models [53] are prevalent in Knowledge Extraction (KE). Other studies incorporated KE methods include relevance detection from social media messages [54], supervised economic event extraction [55], the use of Open Information Extraction for tuple representation [56], utilization of real-time clustering for event extraction [57], and linguistic-based opinion extractions [58]. While these approaches showcase effectiveness, their dependence on labeled data for training poses challenges in terms of resource requirements and restricts adaptability to specific domains. Some studies underscored the significance of gathering context-dependent stock market information to identify the correlation between financial news and asset prices. In a research effort, Atkins et al. [59] employed Naive Bayes and LDA-derived feature vectors to predict volatility, while authors [60] focused on sentiment analysis-based prediction. However, these approaches tended to overlook the importance of temporality, a gap addressed by recent studies [61-62] that focused on enhancing predictive models. CapTE [62] utilizes a Capsule network to capture semantic and structural information from tweets, while Att-DCNN [63] introduces dilated causal convolution networks with attention to event knowledge embedding, achieving superior performance with 72.23% accuracy on different datasets, including the S&P500 index prediction. These studies collectively underscore the evolving landscape of NLP's transformative potential in financial analysis and prediction, emphasizing the significance of temporal analysis and event extraction which is in limited incorporation from temporal aspects in current models.

In summary, the significant gaps identified in the Urdu literature for NLFF not only highlight the absence of financial news data but also underscore the need for extensive exploration in predicting the stock market using Urdu sentiment analysis, leveraging Urdu text resources, and pre-trained deep learning models. Furthermore, the oversight of the crucial relationship or association between stock entities in contemporary stock price prediction research reveals a gap in understanding the intricate dynamics that influence stock prices [64]. The distinct contextual factors impacting the Pakistan Stock Exchange, separate from the advanced economic financial market, emphasize the need for tailored approaches in financial forecasting [10]. Despite the application of natural language processing in financial forecasting, there is still a need to explore forecasting for decision-making based on financial events' timelines, news classification based on NER, and market sentiment.

## 3. Methodology

The main object of this research is to extract, map, and visualize new events timeline which require a sequence of phases, presented in this section. Figure 3. depicts these phases of the proposed research pipeline.

### 3.1 News acquisition and processing module

The initial phase is the creation of a structured financial news dataset for the two sources, i.e., English, and Urdu languages from multiple news sources. To collect the news, a Python-based web scraping process was employed. The news articles in Urdu are collected from Nawaiwaqt, Jang, BBC, and Dunya from 2022 to 2023. For English, news data encompasses Tribune, Daily Times, Pakistan Today, Business Recorder, and Dawn, spanning from 2020 to mid-2023. The selection of these sources is based on their reputation with the business community in Pakistan and their provision of timely and comprehensive news coverage relevant to the Pakistani stock market. To scrap data systematically, a list of URLs leading to different news categories on each news source is compiled. For example, in the case of Dawn, category URLs included links like https://www.dawn.com/pakistan and https://www.dawn.com/business. A comprehensive set of attributes for each news article includes the headline, publication date, URL of the news article, news source, news category as specified in the source, and news description. To ensure data integrity, duplicate URLs are handled in addition to the standardization of the Date formats to form a consolidated raw news dataset. This data contains noise and inconsistencies; therefore, a series of preprocessing steps are implemented on the collected articles to ensure data quality.

The primary text cleaning steps are performed which involve the removal of rows with null values, HTML tags, unnecessary punctuation, stop-words, and special characters in headlines and news descriptions to eliminate textual distractions. The temporal sequence of the news article from varying sources is also maintained by utilizing the Date feature of the dataset. Text standardization with the conversion of news descriptions and headlines to lowercase is also applied in the case of the English language. The category standardization for the dataset is required due to varied terminologies utilized across news sources to represent the news categories such as 'Business' and 'Business & Finance'. For a unified category labeling of the articles, a comprehensive key mapping strategy is employed to align each news source's categories to a standardized dictionary set. In cases where direct mapping isn't available, articles are categorized under the category 'Others.' This systematic approach ensured coherence in the dataset, facilitating streamlined analysis and filtration of the news. Table 1 represents the cross-lingual samples of the systematically evolved raw news data.

Table 1. Raw news samples from English and Urdu data sources

| Language | English | Urdu |
|---|---|---|
| Date | 2021-06-24 | 2023-03-22 |
| Headline | Withdrawal of GST on agri equipment, apparatus suggested | آئی ایم ایف کے ساتھ معاہدے میں تاخیر، شرح سود میں مزید اضافہ کا امکان |
| Url | https://www.brecorder.com/news/40102555/ withdrawal-of-gst-on-agri-equipment- apparatus-suggested | https://urdu.dunyanews.tv/index.php/ur/ Business/709045 |
| Category | Business | Business |
| News | ISLAMABAD: The sub-committee of the National Assembly Special Committee on Agricultural Products has recommended withdrawal of general sales tax (GST) on agricultural equipment and apparatus. The parliamentary panel met under the chairmanship of....... | اسلام آباد: (دنیا نیوز) بین الاقوامی مالیاتی فنڈ (آئی ایم ایف) کے ساتھ سٹاف لیول معاہدے میں تاخیر سے پاکستان کو مختلف معاشی نقصانات کا سامنا ہے۔ ذرائع کے مطابق آئی ایم ایف کیساتھ معاہدے میں تاخیر کے باعث موجودہ 20 فیصد شرح سود میں مزید اضافہ کا امکان ہے، فیصلہ 4 اپریل کو مانیٹری پالیسی کمیٹی کرے گی۔....... |
| Source | Business Recorder | Dunya |

In addition to the consolidated news dataset, financial data from the period 2020-2023 is obtained from publicly available stock data on the official website of the Pakistan Stock Exchange (PSX[1]). The KSE 100 index and its two key sectors i.e., Oil & Gas, and Cement, with a high market capitalization, are focused in this research. The daily open, high, low, close, and volume data for the period of (2020-2023), for the KSE 100 index and the respective companies within each sector, are selected for analysis. Table 2 and Table 3 provide a snapshot reflecting the stock data of KSE 100 and its two prominent sectors.

---

[1] https://dps.psx.com.pk

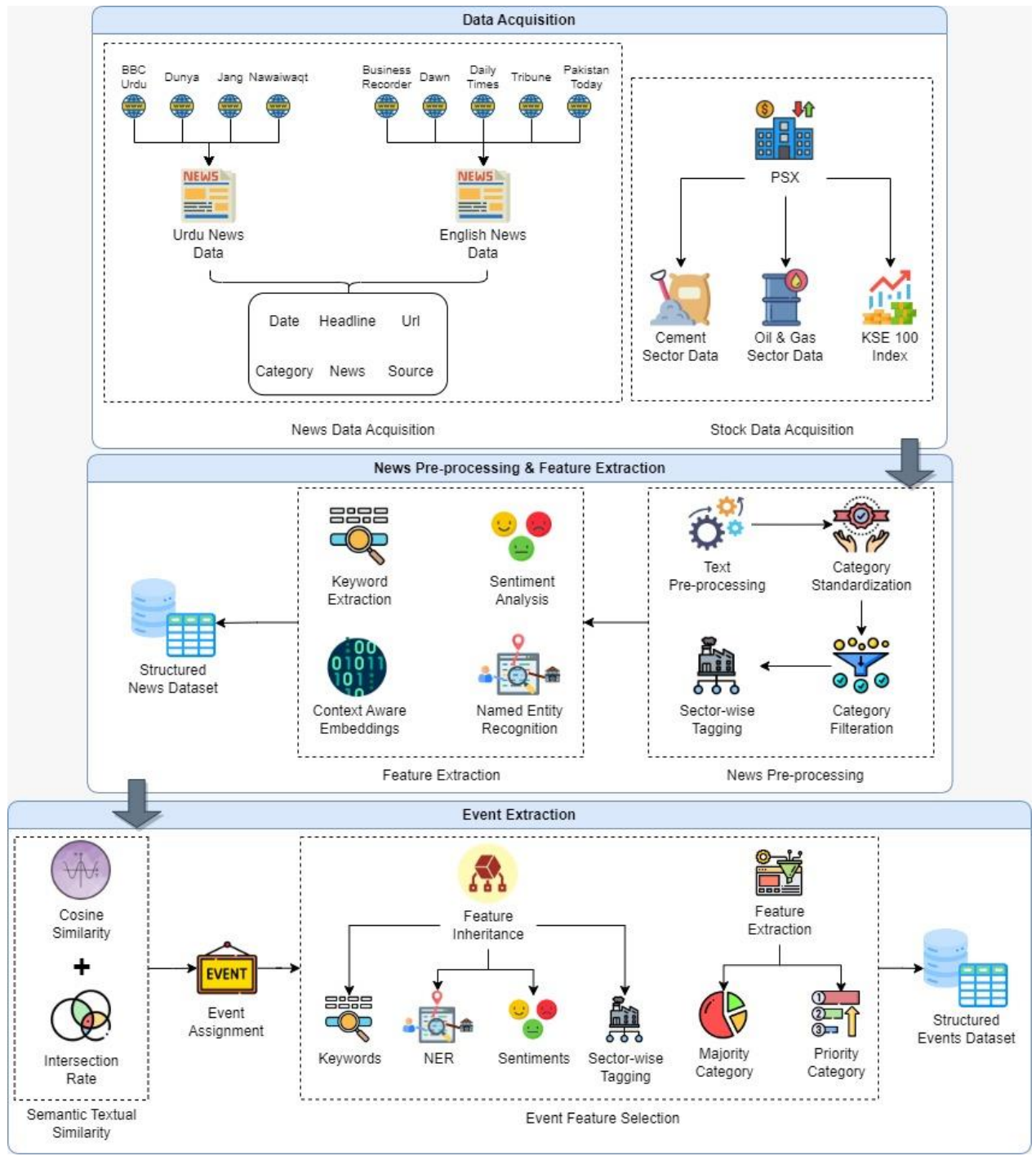

Figure 3. Sequential phases of the proposed event extraction pipeline

                                                                    

Table 2. Sample KSE 100 index stock quotes

| Date | Open | High | Low | Close | Volume | Change% |
|------|------|------|-----|-------|--------|---------|
| 2023-03-24 | 40,376.10 | 40,440.70 | 39,922.89 | 39942.05 | 19472168 | -1.08 |
| 2023-03-27 | 39,942.05 | 39,942.05 | 39,845.06 | 40000.37 | 36310372 | 0.15 |

Table 3. Sample sector-based stock quotes

| Date | Open | High | Low | Close | Volume | Sector | Symbol |
|------|------|------|-----|-------|--------|--------|--------|
| 2023-03-24 | 131.02 | 131.5 | 131.02 | 131.11 | 0.50K | Cement | BWCL |
| 2023-02-01 | 79.1 | 82.65 | 79.01 | 80.22 | 7.75M | Oil & GaS | OGDC |

To compile a comprehensive financial dataset providing insights into the economic and financial landscape of Pakistan, news filtration from the data sources is centered on three retained categories i.e., Pakistan, Business, and World. Furthermore, news articles within the retained categories undergo additional labeling, encompassing single or multiple tags such as sectoral (Oil & Gas, Cement) and economic. A keyword-based labeling mechanism is employed, involving an extensive list of words associated with each tag. If an article contains more than three matching words from the respective keyword lists, the article is subcategorized based on specific tags. The resulting dataset provides insights into economic aspects and sector-specific developments. This approach allows a correlation of the news with the dynamics of the Pakistan Stock Exchange.

As the next phase of the proposed pipeline, feature engineering involves extracting keywords from individual Urdu news articles by employing TF-IDF to identify essential terms with significant meaning within the context of financial information. For English news, both TextRank and TF-IDF are utilized, and the outputs are combined to retain unique keywords for each news article.

The next crucial step is to implement sentiment analysis for financial news data to understand market sentiments and the potential impact on stock performance. For sentiment analysis of our Urdu news dataset, a pre-trained RoBERTa model on Urdu news data is utilized. The model is further fine-tuned with different hyperparameter settings on the sentiment-labeled Urdu dataset[2]. The highest F1Score achieved during the experimentations is 85%. For English news, a popular pre-trained FinBERT model, designed for financial news sentiment classification, is finetuned on the financial review dataset[3] to predict sentiments for English financial news articles achieving an F1Score of 88%. To determine the overall sentiment of a news article, a sentence-level sentiment is calculated, and a majority vote strategy determines the final polarity of each article, categorized as 0 (negative), 1 (neutral), or 2 (positive). The sentiment predictions for our dataset are inspected manually on distinct segments of news, revealing accurate classifications for positive and negative sentiments. However, for few neutral predictions in Urdu news, a closer inspection indicates that news articles exhibit sentiments that align more closely with either the positive or negative category. In the case of English news, few neutral predictions are observed to be potentially positive if classified correctly by the model. This adds room for future refinements by incorporating additional training data to improve the model's performance.

In a continued effort to utilize the NLFF key markers for a comprehensive financial analysis,

---

[2]   https://github.com/mirfan899/Urdu/tree/master
[3]   https://www.kaggle.com/datasets/ankurzing/sentiment-analysis-for-financial-news

the tagging of Named Entity Recognition is applied to the news articles to enhance our understanding of crucial entities in the financial domain. For Urdu news NER tagging, RoBERTa is finetuned on the largest available labeled NER Urdu dataset i.e., MK_PUCIT [13], with three NE types: person, location, and organization. We achieved an F1 Score of 94%, surpassing the state-of-the-art F1 Score reported by the authors in [25] by utilizing MK_PUCIT dataset. News articles from our dataset are tagged with three entity types after tokenization. If a token does not fall into any of the existing named entity types, it is classified as 'other'. After analyzing the results, a few misclassifications were observed for the tags in entities, while some entities were overlooked, lacking tags in any of the three classifications. This highlights the need for further fine-tuning on either a larger dataset or to include new examples to improve the model's entity tagging accuracy. Leveraging Flair's pre-trained capabilities on ner-english, we focused on acquiring persons and organizations tags for English news articles, while encompassing a third class for miscellaneous entities. Manual inspections have revealed promising results, particularly in accurately identifying individuals from Pakistan, showcasing Flair's effectiveness in capturing details within the context of financial news.

### 3.2 Event extraction module

Event extraction from news articles plays a pivotal role due to the utilization of data from distinct news sources in the proposed research pipeline. To proceed with the structured financial news dataset, embeddings are calculated to capture the semantic features and contextual information from the news articles using pre-trained RoBERTa with a 768-dimension dataset. As a next step, a strategy is formulated to identify and extract potential events from the news sources.

It is noteworthy that, when a significant financial event occurs on a specific date, it prompts the release of news articles by all sources. For instance, consider the event "Pakistan's Government announcement of new economic policies affecting the stock market". Each source is likely to feature an article covering this event within their financial sections. This allows grouping all articles from diverse sources under a unified event in a systematic approach. The proposed event identification approach computes a weighted sum of the cosine similarity score of embeddings and the intersection rate of overlapping news articles, providing a robust measure of the semantic similarity of articles for event extraction.

#### 3.2.1 Event extraction approaches

The event extraction approach is spanned over two periods i.e., "1-Day Event Extraction" and "5-Day Event Extraction". The prior approach involves identifying events among news articles that share the same publication date, leveraging the similarities between these articles within the same day as a financial event occurs. In contrast, the "5-Day Event Extraction" approach utilizes a more extended temporal window for extracting events that evolve over several days.

Let $D = [d_0, d_{o+1}, d_{o+2}, \ldots, d_L]$ be a list of all the unique consecutive dates of the structured news dataset SN, where $d_o$ is the oldest date and $d_L$ is the latest date.

Let $W = [w_1, w_2, \ldots, w_n]$ be the list of 5 day rolling window sizes for D, where $w_1 = [d_0, d_{o+1}, d_{o+2}, d_{o+3}, d_{o+4}]$, $w_2 = [d_{o+1}, d_{o+2}, d_{o+3}, d_{o+4}, d_{o+5}]$, $w_n = [d_{n-4}, d_{n-3}, d_{n-2}, d_{n-1}, d_n]$, and n is the number of windows created based on the total number of consecutive dates in $D$:

$$n = |D| - (Window\ Size) + 1 \qquad\qquad \text{Eq (1)}$$

where $|D|$ is the length of $D$ and Window Size = 5. If |D| = 31 we will have 27 windows.

$\forall\ w \in W\ let\ df(w)$ be the subset of SN where the dataset df contains the information of the news articles only for the window $w$.

$$df(w) = SN('Date' == w) \qquad \text{Eq (2)}$$

Let E(df) define the function for event extraction for the dataset df.

$$SE_2 = \sum_{i=w_1}^{w_n} E(df(i)) \qquad \text{Eq (3)}$$

where $SE_2$ is the structured event dataset for 5-Day Event Extraction.

### 3.2.2 Semantic Similarity Calculation

The first mechanism involves a calculation of the cosine similarity score for the embeddings of the news articles in the dataset, enabling the determination of their semantic similarity. Cosine similarity measures the cosine of the angle $\theta$ between two non-zero vectors in a multi-dimensional space. The mathematical formula for cosine similarity between two vectors A and B can be expressed in Equation 4. as follows:

$$Cosine\ Similarity = cos\theta = \frac{A \cdot B}{\|A\|\|B\|} \qquad \text{Eq (4)}$$

where: $A \cdot B$ represents the dot product of vectors A and B. $\|A\|$ represents the magnitude (Euclidean norm) of vector A. $\|B\|$ represents the magnitude (Euclidean norm) of vector B. Higher cosine similarity values indicate greater similarity between the vectors, where 1 represents identical vectors, and 0 represents orthogonal vectors.

The second mechanism involves the utilization of the intersection rate approach to quantify the overlap in the content between articles. This rate is computed by comparing the words shared between two articles, relative to the total words in the smaller article. The mathematical equation for calculating the intersection rate can be represented in Equation 5. as follows:

$$Intersection\ Rate = \frac{|wl_1 \cap wl_2|}{|wl_s|} \qquad \text{Eq (5)}$$

$$\text{where:}\ wl_s = min\ (wl_1, wl_2)$$

$wl_1$ and $wl_2$ are the lists of words of two articles and $|wl_1 \cap wl_2|$ represents the cardinality (number of elements) in the intersection of the two lists, which is the count of common elements between them. Whereas $|wl_s|$ represents the cardinality of the smaller of the two lists, ensuring that the intersection rate is calculated relative to the size of the smaller list.

### 3.2.3 Threshold optimization for similarity score

Based on the relative importance of the intersection rate and the cosine similarity a combined similarity score is computed. Several experiments are performed to acquire an optimized threshold for the final similarity score for event assignment. A similarity score greater than 0.8, coupled with an intersection rate exceeding 0.5 between any two news articles, serves as the optimized threshold criterion. Finally, the articles satisfying these conditions are effectively identified as part of the same event. Equation 6. represents the optimized threshold values for each:

$$Similarity\ Score = (Cosine\ Similarity \times 0.8) + (Intersection\ Rate \times 0.2) \qquad \text{Eq (6)}$$

### 3.2.4 Event feature extraction

The event assignment process is further refined for event selection in the events timeline generation. This involves incorporating feature inheritance from the group of event-relevant news articles, to establish the contextual foundation of the identified events. Furthermore, for insightful analysis of the correlated news events in the stock market and its sectors, the mapping of the selected events on the timeline also includes the categorization of events based on the most prevalent category among the constituent articles. The attributes contributing to this categorization involve the Majority Category, Priority Category, and their associated purities. The Majority Category signifies the category to which most news articles within an event belong. Priority levels for determining the event category are assigned with category priorities as follows: {Business (1), Pakistan (2), and World (3)}. Additionally, purity calculations offer a quantifiable metric, allowing us to assess the extent of dominance exhibited by the selected category within a given event. A purity of 1.0 indicates complete dominance of the category in event-related news, while a threshold of >= 0.5 is acceptable in our case, ensuring a better representation of the category in the event-related news articles. This approach not only helps understand the predominant themes within the news dataset but also provides insights from diverse sources. Thus, the comprehensive approach empowers us to delve into the intricate relationship between news content, event extraction, and their potential impact on stock sectors.

Our proposed research pipeline enables the visualization of the correlation between news and stock data, leveraging the persistent features we have calculated and maintained throughout the analysis.

## 4. Results and Discussion

In our experimentation, optimization, and fine-tuning of the models involved conducting necessary experiments on an RTX 3060 GPU with a memory size of 12GB. For news scrapping, Python libraries including Selenium, Requests, and BeautifulSoup are utilized. The Urdu news dataset comprises of 53,078 news articles, while the English news dataset consists of 127,912 news articles. Due to the complexity and scale of the data, the news articles were split into monthly datasets. Each monthly dataset contains aggregated news from all sources, facilitating efficient processing for subsequent tasks in the proposed pipeline. Various NLP tasks are executed using the Python version (3.10.12) and advanced libraries built on Keras and PyTorch. Table 4 outlines the specifics of the experimental settings utilized in fine-tuning the models for essential NLP tasks along with the evaluation metric where applicable.

Table 4. Specifics of the experimental environment

| Language | Task | Model | Hyperparameters | F1 Score |
|---|---|---|---|---|
| *Urdu* | Sentiment Analysis | urduhack/roberta-urdu-small | optimizer: Adam<br>learning rate: 1e-5<br>clipnorm: 1<br>epochs: 7 | **0.85** |
| | NER | | num_train_epochs: 10<br>early_stopping_patience: 3<br>learning_rate: 2e-5 weight_decay: 0.01 per_device_train_batch_size: 64 per_device_eval_batch_size: 64<br>num_of_labels: 4 | **0.94** |
| | Embeddings | RoBERTa-base | max_length: 512<br>output_vector_length: 768 | - |
| *English* | Sentiment Analysis | ProsusAI/finbert | optimizer: AdamW<br>learning rate: 2e-5<br>loss_function: CrossEntropyLoss<br>batch size: 16<br>num_of_labels: 3<br>epochs: 4 | **0.88** |
| | Embeddings | RoBERTa-base | max_length: 512<br>output_vector_length: 768 | - |

The visualization results of the news embeddings demonstrate the capacity and versatility of the RoBERTa model in encoding rich contextual information for both languages. Figure 4 represents the effective semantic representation of the news embeddings concerning the categories from our dataset for May 2023. The proximity of the embeddings within their respective color sections highlights the model's effectiveness in capturing semantic similarities and grouping similar articles. The prevalence of news distribution in 'Pakistan' as a main category is prominent in the t-SNE visualization, suggesting that a significant number of events captured in the news articles are centered around or impact Pakistan. This insight is valuable for understanding the regional dynamics in Pakistan, providing a comprehensive overview of the news landscape.

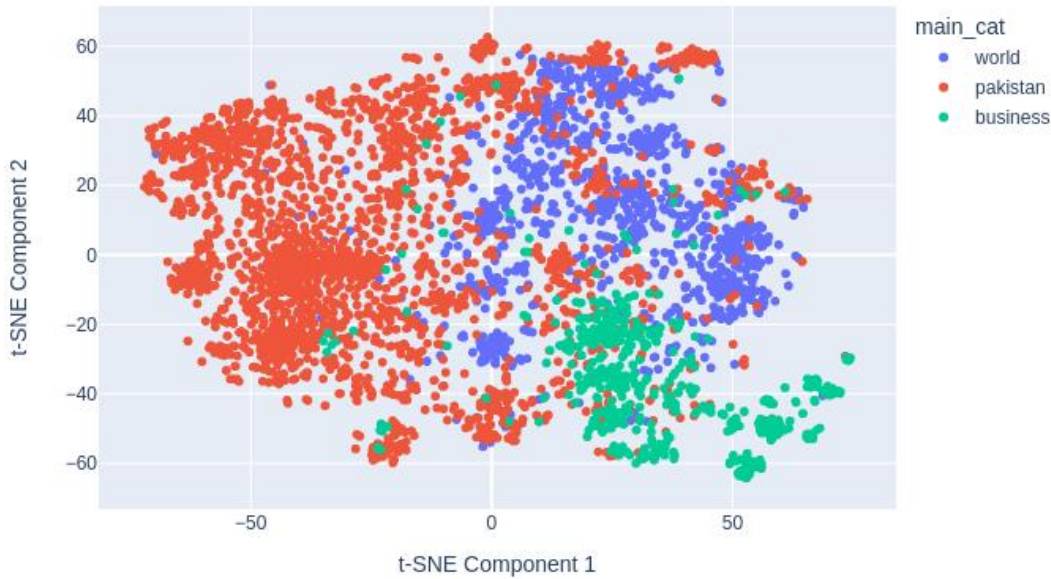

Figure 4. t-SNE visualization of RoBERTa embeddings for news categories clusters

The labeling of the financial news dataset, encompassing subcategories from economic and stock market sectors, provides useful insights into the cross-lingual sources. The correlation of the news events with the dynamics of the Pakistan Stock Exchange is shown in Table 5. A prevalent focus on the economic sector in most articles is observed as compared to specific stock industries. Extracting and filtering events from each labeled sector for representation in the financial event timeline reveals that the business community in Pakistan exhibits a notable interest in English news sources. It can be inferred from the statistics that this preference may stem from the perceived quality and trustworthiness associated with such sources.

Table 5. News and events statistics from the structured cross-lingual datasets for the year 2022

| Language | Sector | No. of news articles | No. of events extracted |
|---|---|---|---|
| Urdu | Oil & Gas | 994 | 109 |
| | Cement | 358 | 34 |
| | Economic | 19,197 | **1632** |
| English | Oil & Gas | 13024 | 3512 |
| | Cement | 685 | 279 |

| | Economic | 20453 | **5059** |
|---|---|---|---|

A sample of two extracted events is presented in Table 6. The initial event pertains to the cement sector, involving the contribution of three news articles from two distinct sources. In contrast, the second event assignment is represented by four articles originating from two Urdu news sources. For each event instance, the similarity score, majority category, and sentiment are calculated by following the procedure as outlined in sections 3.2.3 and 3.2.4.

Table 6. Cross-lingual events sample from the structured financial dataset

| Event Name | Date | Sources | Articles | Description | Majority Category | Sector Category | Sentiment | Similarity score |
|---|---|---|---|---|---|---|---|---|
| Cement sales dip 16.6% in Jan | 2022-02-03 | Dawn, Tribune | 3 | KARACHI: The local cement sales dropped 16 percent year-on-year to 3.4 million tonnes followed by a 21pc fall in exports to 551,006 tonnes in January …… | Business | Cement | 0 | 0.9 |
| سٹیٹ بینک آف پاکستان کے زرمبادلہ ذخائر میں 28 کروڑ ڈالر اضافہ | 2023-03-24 | Jang, Dunya | 4 | ملکی زرمبادلہ ذخائر میں مسلسل آٹھویں بفتہ اضافہ ہوا ہے، جس کے نتیج میں زرمبادلہ ذخائر 10 ارب ڈالر کی سطح عبور کرگئے۔۔۔ | Business | Economic | 2 | 0.85 |

Figure 5 shows the fortnightly financial timeline with respective event sentiment polarities on the KSE 100 trend. This assists investors in foreseeing potential market movements and understanding historical correlations between news and financial performance, enabling them to formulate and adapt investment strategies effectively. This is also evident from the figure which depicts that the negative news articles are much greater in number as compared to the positive ones and hence we can see a decline in the market trend.

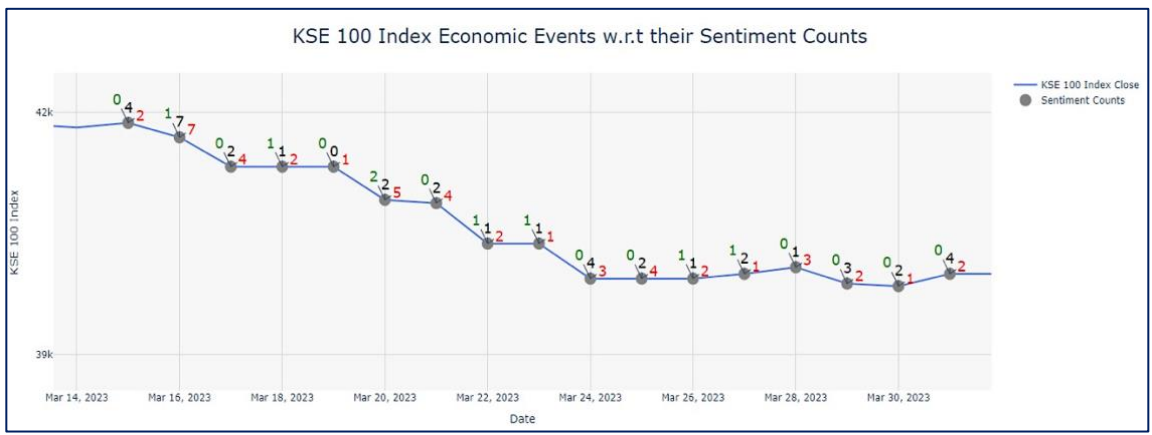

Figure 5. Financial events trend with respective sentiment count on KSE 100

Figure 6 shows the fortnightly financial timeline with respective event sentiment polarities from English sources on the Oil & Gas sector trend. It becomes apparent that the polarity of events directly influences the

 

trend movement. When both positive and negative sentiments are aligned, a notable peak in the trend emerges. This phenomenon suggests that the convergence of opposing sentiments may create heightened market volatility or indicate a potential turning point. Further analysis is needed to explore the underlying factors contributing to this distinctive pattern. In addition to the summary statistics of the sentiments of the event, the interface of our model allows users to explore additional insights for any business day. This encompasses keywords, event descriptions, and entity extraction of each event.

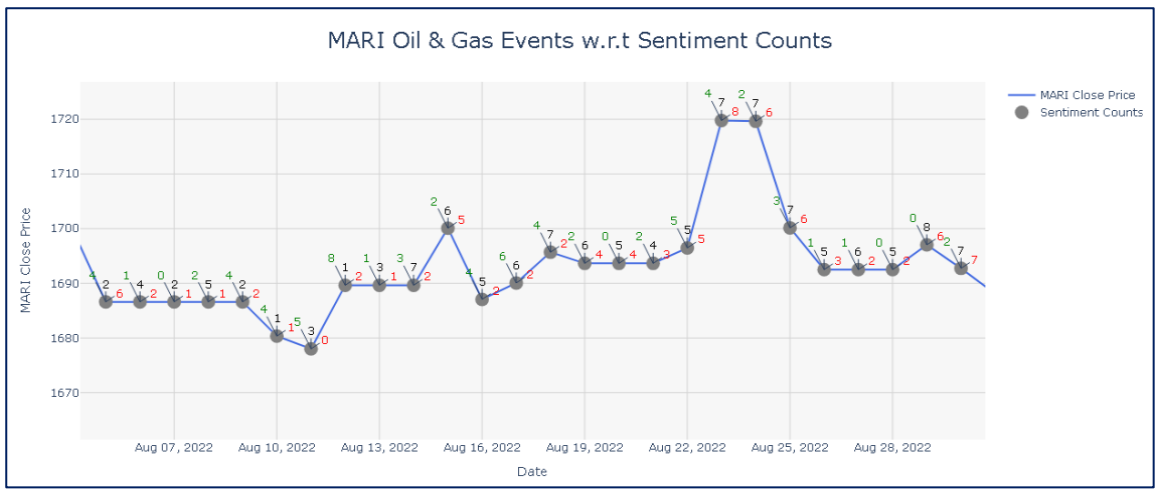

Figure 6. Financial events trend with respective sentiment count on Oil & Gas sector

The efficacy of the financial events timeline in strategically extracting and mapping events is presented in Figure 7. During this time interval, the confidence of investors was boosted by the investment news in different business sectors, relaxation in the stock investor's eligibility criteria, and the increase in foreign remittance. These factors attributed to a significant increase in the stock momentum and resulted in a positive market trend. On the contrary, the prevailing sentiment of most events, as presented in Figure 8, is negative, contributing to a noticeable downtrend in the KSE 100 index. Significant events that are observed in the trend are fiscal deficit, a decline in State Bank reserves, an increase in interest rate, and settlement issues with the IMF for loan packages. The instantaneous occurrences of these events in a short span exerted a substantial negative impact on the stock market trend in the near future. It is noteworthy that our proposed model is able to signify key events and their association related to the stock movement and hence provides an instrument for highlighting vital information to unlock rich insights from vast unstructured business-related news data.

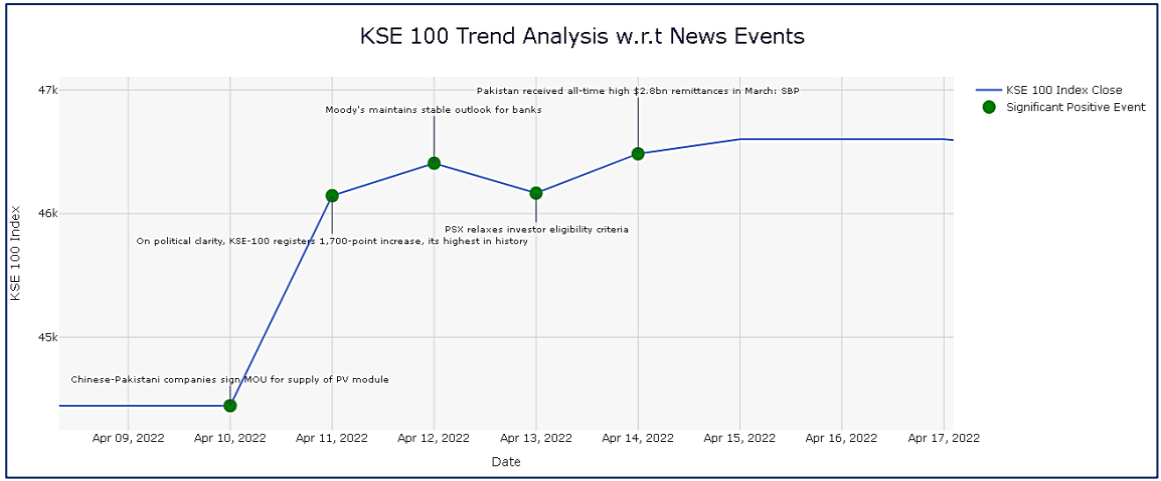

Figure 7. Visualization of key news events timeline aligned with KSE 100 bullish trend.

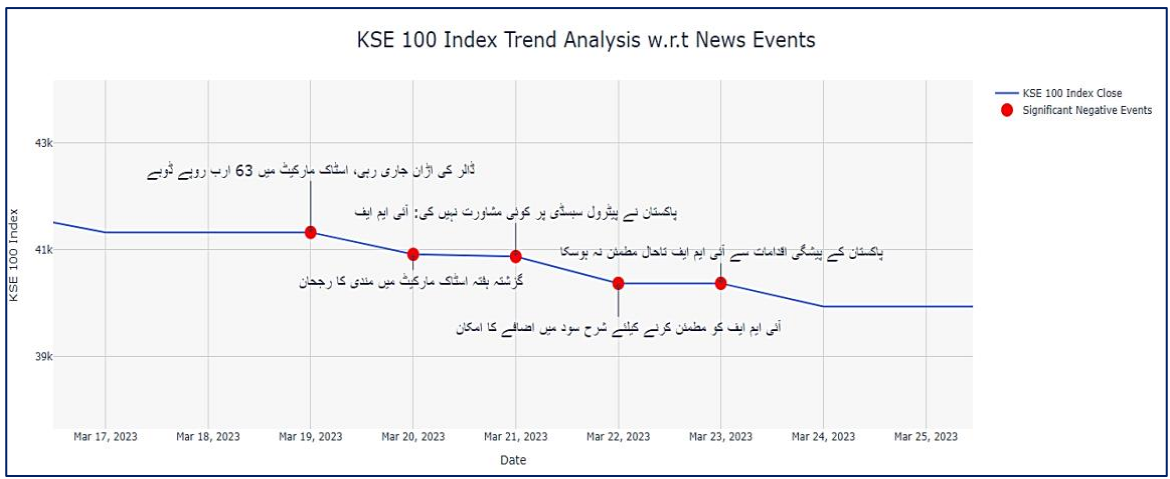

Figure 8.  Visualization of key news events timeline aligned with KSE 100 bearish trend.

## 5.  Conclusion and Future Direction

In this study, we proposed a systematic cross-lingual NLFF pipeline, leveraging key markers in the domain. These markers encompass semantic similarity, sentiment analysis, and named entity recognition to generate a financial events timeline, extracted from the news sources. Keeping in view the economic landscape of Pakistan, we have curated a structured financial dataset in two language sources, Urdu, and English. With news articles embeddings along with a similarity threshold optimization strategy, we successfully identified and extracted events, candidate for financial timeline generation. Financial timeline visualizations were produced to anticipate the stock market trend and assist the investors in future stock investments. We successfully correlated stock price movements with cross-linguistic news sentiments to a significant degree.

Several research avenues can be investigated as a future of this research effort. It is evident that stock prices are not influenced by all business-related news and hence the identification of more relevant events for stock prices will be an interesting area to explore. One possible direction is to understand the association between news sentiments and stock prices by investigating the potential impact of positivity or negativity in news articles and events. Furthermore, there is room for improvement in Urdu sentiment analysis, specifically in distinguishing negatives that are frequently misassigned as neutral. This is crucial for accurate event sentiment assignment and effective stock trend analysis from Urdu news sources. The identification of the relationship between stock entities and events to forecast future stock trends is another dimension that can be explored by using a knowledge graph. Knowledge graphs empower financial analysts to facilitate the identification of patterns and trends within complex datasets providing rich data insights and capturing vital information. Furthermore, a lower ratio of news events, pertaining to specific stock sectors as compared to economic events, was observed in most of the cases. This opens avenues for further exploration in analyzing sector-specific trends for unique opportunities within specific sectors. We plan to implement a comprehensive financial forecasting system integrating technical indicators derived from historical stock data with a diverse range of economic indicators. This approach will not only augment the overall forecasting results but will also allow for a thorough evaluation of the financial event timeline generated through our study. Thus, enabling the anticipation of market movements and the personalized generation of investment decisions based on investor's preferences.

**Acknowledgement**

This research is funded by the Higher Education Commission's National Research Program for Universities (NRPU) under project number 20-15756 and titled "Generating Plausible Counterfactual Explanations using Transformers in Natural Language-based Financial Forecasting (NLFF)". We are deeply grateful for HEC-NRPU financial support, which allowed us to conduct the experiments and analysis presented in this paper. This support has facilitated collaboration and knowledge-sharing among our research team and Falki Capital (Pvt) Ltd. (Equities brokerage house).

**Source code and dataset availability statement**

For the reproducibility of our results, the source code is available on the GitHub repository at https://github.com/Seemab05/NRPU. Raw data were collected from online news sources. Derived data supporting the findings of this study are available from the corresponding author, Seemab Latif, on request.